\documentstyle [12pt] {article}

\catcode`@=12
\begin{document}
\thispagestyle{empty}
\begin{flushright} UCRHEP-T146\\July 1995\\ \end{flushright} \vspace{0.5in}
\begin{center} {\Large \bf Decay of Z into Two Light Higgs Bosons\\}
\vspace{1.5in} {\bf T. V. Duong, E. Keith, Ernest Ma\\} {\sl Department of
Physics, University of California, Riverside, California 92521\\}
\vspace{0.1in}
{\bf Hisashi Kikuchi\\} {\sl Ohu University, Koriyama, Fukushima 963, Japan\\}
\vspace{1.5in}
\end{center}
\begin{abstract}\ If the standard electroweak gauge model is extended to
include
two or more Higgs doublets, there may be a neutral Higgs boson $h$ which is
light (with a mass of say 10 GeV) but the $hZZ$ coupling is suppressed so that
it has so far escaped experimental detection. However, the effective $hhZZ$
coupling is generally unsuppressed, hence the decay of Z into two light Higgs
bosons plus a fermion-antifermion pair may have an observable branching
fraction, especially if $h$ decays invisibly as for example in the recently
proposed doublet Majoron model. \end{abstract}

\newpage
\baselineskip 24pt
\section{Introduction}

In the standard model, the Z boson may decay into the Higgs boson H and a
fermion-antifermion pair. From the absence of such events, it has been deduced
that $m_H$ is greater than about 60 GeV\cite{1}. On the other hand, in
extensions of the standard model with two or more scalar doublets, there may be
a neutral Higgs boson $h$ with $m_h < 60$ GeV which has escaped experimental
detection so far because the $hZZ$ coupling is suppressed\cite{2}. This
situation can be the natural consequence of a particular theoretical model, an
example\cite{3} of which will be discussed in this paper.

Our main observation is that the effective $hhZZ$ coupling is generally
unsuppressed, hence the decay of Z into two $h$'s plus a fermion-antifermion
pair may have an observable branching fraction. Note that the decay $Z
\rightarrow hh$ is strictly forbidden because of angular-momentum conservation
and Bose statistics. In Section 2 we formulate our analysis in the context of
two Higgs doublets. We obtain the condition for the possible existence of a
light $h$ which would not conflict with present data. In Section 3 we identify
all the contributions to the effective $hhZZ$ coupling. We show that the
gauge-sector contributions alone are probably not large enough for our proposed
process to be observable, but the addition of scalar-sector contributions may
make it so. In Section 4 we focus on the recently proposed doublet Majoron
model\cite{3} where $h$ decays invisibly and discuss a related issue. Finally
in
Section 5 there are some concluding remarks.

\section{Two Higgs Doublets}

Consider the following Higgs potential $V$ for two $SU(2) \times U(1)$ scalar
doublets $\Phi_{1,2} = (\phi^+_{1,2}, \phi^0_{1,2})$: \begin{eqnarray} V &=&
\mu_1^2 \Phi_1^\dagger \Phi_1 + \mu_2^2 \Phi_2^\dagger \Phi_2 + \mu_{12}^2
(\Phi_1^\dagger \Phi_2 + \Phi_2^\dagger \Phi_1) \nonumber \\ &+& {1 \over 2}
\lambda_1 (\Phi_1^\dagger \Phi_1)^2 + {1 \over 2} \lambda_2 (\Phi_2^\dagger
\Phi_2)^2 + \lambda_3 (\Phi_1^\dagger \Phi_1) (\Phi_2^\dagger \Phi_2) \nonumber
\\ &+& \lambda_4 (\Phi_1^\dagger \Phi_2) (\Phi_2^\dagger \Phi_1) + {1 \over 2}
\lambda_5 (\Phi_1^\dagger \Phi_2)^2 + {1 \over 2} \lambda_5^* (\Phi_2^\dagger
\Phi_1)^2.
\end{eqnarray} Assume $\lambda_5$ to be real for simplicity. Define $\tan \beta
\equiv v_2/v_1$ as is customary, where $v_{1,2} = \langle \phi^0_{1,2} \rangle$
are the usual two nonzero vacuum expectation values. The charged Higgs boson is
then given by
\begin{equation} H^\pm = \sin \beta \phi_1^\pm - \cos \beta \phi_2^\pm,
\end{equation}
\begin{equation} m^2_{H^\pm} = - \mu_{12}^2 (\tan \beta + \cot \beta) -
(\lambda_4 + \lambda_5) (v_1^2 + v_2^2); \end{equation} the pseudoscalar
neutral
Higgs boson is given by \begin{equation} A = \sqrt 2 (\sin \beta {\rm Im}
\phi_1^0 - \cos \beta {\rm Im}
\phi_2^0), \end{equation}
\begin{equation} m_A^2 = - \mu_{12}^2 (\tan \beta + \cot \beta) - 2 \lambda_5
(v_1^2 + v_2^2); \end{equation} and the two scalar neutral Higgs bosons $\sqrt
2
{\rm Re} \phi^0_{1,2}$ have the following mass-squared matrix: \begin{equation}
{\cal M}^2 = \left[ \begin{array} {c@{\quad}c} -\mu_{12}^2 \tan \beta + 2
\lambda_1 v_1^2 & \mu_{12}^2 + 2 (\lambda_3 + \lambda_4 + \lambda_5) v_1 v_2 \\
\mu_{12}^2 + 2 (\lambda_3 + \lambda_4 + \lambda_5) v_1 v_2 & -\mu_{12}^2 \cot
\beta + 2 \lambda_2 v_2^2 \end{array} \right]. \end{equation} Let us rotate to
the basis of $H = \sqrt 2 (\cos \beta {\rm Re} \phi_1^0 + \sin \beta {\rm Re}
\phi_2^0)$ which couples singly to the Z, and $H' =
\sqrt 2 (\sin \beta {\rm Re} \phi_1^0 - \cos \beta {\rm Re} \phi_2^0)$ which
does not. Then \begin{equation} {\cal M}_{11}^2 = 2 [\lambda_1 \cos^4 \beta +
\lambda_2
\sin^4 \beta + 2(\lambda_3 + \lambda_4 + \lambda_5) \sin^2 \beta \cos^2 \beta]
(v_1^2 + v_2^2),
\end{equation}
\begin{equation} {\cal M}_{22}^2 = -\mu_{12}^2 (\tan \beta + \cot \beta) +
2(\lambda_1 + \lambda_2 - 2\lambda_3 - 2\lambda_4 - 2\lambda_5) \sin^2 \beta
\cos^2 \beta (v_1^2 + v_2^2),
\end{equation}
\begin{equation} {\cal M}_{12}^2 = {\cal M}_{21}^2 = -\sin 2 \beta [\lambda_1
\cos^2 \beta - \lambda_2 \sin^2 \beta - (\lambda_3 + \lambda_4 + \lambda_5)
\cos
2 \beta] (v_1^2 + v_2^2).
\end{equation} Hence $H'$ may well have escaped experimental detection if its
mixing with $H$ is small. This is clearly the case if $|\sin 2 \beta| << 1$.
Without loss of generality, let us consider $v_1 << v_2$, {\it i.e.} $\sin
\beta
\simeq 1$ and $\cos \beta \simeq 0$; and assume ${\cal M}_{22}^2$ to be small.
Then $h \simeq H'$ and
\begin{eqnarray} m_h^2 &\simeq& {\cal M}_{22}^2 - {\cal M}_{21}^2 {\cal
M}_{12}^2 / {\cal M}_{11}^2 \nonumber \\ &\simeq& -\mu_{12}^2 \tan \beta + 2
\lambda_1 v_1^2 \left[ 1 - {{(\lambda_3 + \lambda_4 + \lambda_5)^2} \over
{\lambda_1 \lambda_2}} \right].
\end{eqnarray} The second term in the above is naturally small, but the first
term is not, unless $\mu_{12}^2$ is fine-tuned to be of order $v_1^3/v_2$. This
means that unless $\mu_{12}^2 = 0$ from a symmetry requirement of the model, it
is not likely that a light neutral Higgs boson has so far escaped experimental
detection. Furthermore, since the $hAZ$ coupling is unsuppressed in this limit,
the nonobservation of $Z \rightarrow hA$ at the $e^+e^-$ collider LEP at CERN
means that $m_h + m_A > M_Z$. Comparing Eq.~(10) with Eq.~(5), we see that
$\lambda_5 \neq 0$ is another necessary condition. By the same token, the
nonobservation of $Z \rightarrow H^+H^-$ requires $\lambda_4 + \lambda_5
\neq 0$.

In the minimal supersymmetric standard model (MSSM), \begin{equation} \lambda_1
= \lambda_2 = -\lambda_3 - \lambda_4 = {1 \over 4} (g_1^2 + g_2^2), ~~~
\lambda_5 = 0.
\end{equation} Hence $m_h \simeq m_A \simeq -\mu_{12}^2 \tan \beta$ for large
$\tan \beta$ and the absence of $Z \rightarrow hA$ events implies $m_h >
M_Z/2$.
This means that Z decay into two $h$'s would not be possible kinematically and
our proposal cannot be realized in the MSSM. On the other hand, a natural model
with $\mu_{12}^2 = 0$ and $\lambda_5 \neq 0$ does exist. It is the recently
proposed doublet Majoron model\cite{3}, details of which will be discussed
later
in Section 4.

\section{Effective hhZZ Coupling}

We assume $h$ to be light and $A$ to be heavy so that $Z \rightarrow hA$ is
kinematically forbidden, whereas $Z \rightarrow hh \bar {f} f$ is allowed, $f$
being either a quark or a lepton. The contributing diagrams are given in
Fig.~1.
(We have assumed that Yukawa couplings of $f$ to $h$ are negligible. Actually
there is an important exception if the $b$ quark has a large Yukawa coupling to
both $h$ and $A$. In that case, the $Z \rightarrow \bar {b} b$ rate may be
enhanced to explain the $R_b$ excess observed at LEP, but then $Z \rightarrow
\bar {b} b + h(A)$ should become observable\cite{4} with a branching fraction
of
order 10$^{-4}$.) We can eliminate diagram (c) because it is negligible for
large $\tan \beta$. As for diagrams (b) and (d), they may be suppressed for
large values of $m_H$ and $m_A$ respectively. The only model-independent
contribution is that of diagram (a). The fundamental $hh Z^\mu Z^\nu$ coupling
is always unsuppressed, with Feynman rule given by $ig^2 g^{\mu\nu}/2\cos^2
\theta_W$, same as that for $HH Z^\mu Z^\nu$. Let the $Z \bar {f} f$ coupling
be
given by $(g/\cos \theta_W) Z^\mu \bar {f} \gamma_\mu (a + b \gamma_5) f$,
where
$a = I_{3L}/2 - Q \sin^2 \theta_W$ and $b = -I_{3L}/2$, $I_{3L}$ and $Q$ being
the weak-isospin projection of $f_L$ and electric charge of $f$ respectively.
Integrating out the momenta of the two $h$'s, we obtain the differential decay
rate of $Z \rightarrow hh \bar {f} f$ as a function of the energies $E_3$ and
$E_4$ of the fermion-antifermion pair, and the angle $\theta_{34}$ between
them:
\begin{equation} {{d\Gamma} \over {d E_3 d E_4 d \cos \theta_{34}}} = {{g^6
(a^2
+ b^2)} \over {3072 \pi^5 \cos^6 \theta_W}} {{E_3^2 E_4^2 (3 - \cos
\theta_{34})
\sqrt {1-\Delta}} \over {M_Z [M_Z^2 - 2 E_3 E_4 (1-\cos \theta_{34})]^2}},
\end{equation} where
\begin{equation}
\Delta = {{4m_h^2} \over {M_Z(M_Z - 2E_3 - 2E_4) + 2E_3 E_4 (1-\cos
\theta_{34})}}.
\end{equation} We have assumed in the above that $m_f$ can be neglected. The
kinematical constraints are
\begin{equation} 0 < E_3 < {1 \over 2} M_Z \left[ 1 - {{4m_h^2} \over M_Z^2}
\right], ~~~ -1 < \cos \theta_{34} < 1,
\end{equation}
\begin{equation} 0 < E_4 < {1 \over 2} M_Z \left[ {{M_Z - 2E_3 - 4m_h^2/M_z}
\over {M_Z - E_3 (1-\cos \theta_{34})}} \right]. \end{equation} We integrate
the
above numerically and sum over all fermion species, {\it i.e.} the quarks $u$,
$d$, $s$, $c$, $b$, and the three families of leptons $\nu_i$, $l_i$. The
resulting total branching fraction for $Z \rightarrow hh
\bar {f} f$ as a function of $m_h$ is shown in Fig.~2. Using the experimental Z
width of 2.5 GeV, the branching fraction is thus only about $1.8 \times
10^{-8}$
for $m_h = 10$ GeV. With the accummulation of about $8 \times 10^6$ Z decays at
LEP up to the end of 1993, this amounts to only 0.15 event. This means that if
only diagram (a) is important, we do not expect this decay to be readily
observable. The contribution of diagram (d) is also small, {\it i.e.}
comparable
in magnitude to that of diagram (a), because $m_A$ must be greater than $M_Z -
m_h$ and there are no other adjustable parameters.

The contribution of diagram (b) depends on the $hhH$ coupling with Feynman rule
given by $-i \sqrt 2 (\lambda_3 + \lambda_4 + \lambda_5) v_2$ for large $\tan
\beta$. Now $m_H^2 \simeq 2 \lambda_2 v_2^2$ in this limit, hence the effective
$hhZZ$ contribution in diagram (b) is given by \begin{equation} {{ig^2 g^{\mu
\nu}} \over {2 \cos^2 \theta_W}} {{2(\lambda_3 + \lambda_4 + \lambda_5) v_2^2}
\over {(k_1 + k_2)^2 - 2 \lambda_2 v_2^2}}, \end{equation} where $k_{1,2}$ are
the four-momenta of the two $h$'s. An enhancement of the $Z \rightarrow hh \bar
{f} f$ rate is thus possible if the ratio $(\lambda_3 + \lambda_4 +
\lambda_5)/\lambda_2$ is large enough. Further enhancement occurs if $m_H$ is
not much greater than its experimental lower bound of about 60 GeV. Assuming
that diagram (b) dominates, we show in Fig.~3 the branching fraction $B$ of $Z
\rightarrow hh \bar {f} f$ for various values of $|\lambda_3 +
\lambda_4 + \lambda_5|$, $m_h$, and $m_H$. We note that $|\lambda_3 + \lambda_4
+ \lambda_5|$ should not be too large, because $\lambda_1 - (\lambda_3 +
\lambda_4 + \lambda_5)^2/\lambda_2$ is constrained by the smallness of $m_h^2$
to be at most of order unity and we want to avoid having to fine-tune
$\lambda_1$. We see from Fig.~3 that there is a significant region in parameter
space with $B > 10^{-6}$, in which case our proposed process may in fact become
observable.

So far we have not specified how the scalar doublets couple to quarks and
leptons. If we consider the usual case of $\Phi_1 (\Phi_2)$ coupling to
down(up)-type quarks and charged(neutral) leptons, then $h$ decays mainly into
$b \bar b$ if kinematically allowed. The final state of our process would then
contain $b \bar {b} b \bar {b}$ + another fermion-antifermion pair. The
background to this from second-order QCD (quantum chromodynamics) radiative
corrections has not been calculated, but we estimate it to be of order
10$^{-5}$
to 10$^{-6}$. This would make it very difficult for the observation of $Z
\rightarrow hh \bar {f} f$ unless its branching fraction is much greater than
10$^{-6}$. On the other hand, in some theoretical models, $h$ decays
invisibly\cite{5}. It may appear at first sight that this would be more
difficult to detect experimentally\cite{6}. Actually the opposite is true at
LEP
because the well-defined missing energy provides a signature with essentially
no
background and better limits are possible if $h$ decays invisibly than
otherwise\cite{2}. In the following section we will discuss the scalar sector
of
the recently proposed doublet Majoron model which has all the desired
properties
for the possible observation of $Z \rightarrow hh \bar {f} f$, {\it i.e.}
$\mu_{12}^2 = 0$, $\lambda_5 \neq 0$, and $h$ decays invisibly to two massless
Goldstone bosons.

\section{Doublet Majoron Model}

The scalar sector of the doublet Majoron model\cite{3} consists of three $SU(2)
\times U(1)$ doublets with lepton number assignments $L = 0$, 1 and $-1$. The
Lagrangian is assumed to conserve $L$, hence terms of the form $\Phi_i^\dagger
\Phi_j$ for $i \neq j$ are not allowed. The Higgs potential is also assumed to
be symmetric under the interchange of the two scalar doublets with $L = \pm 1$.
As all three doublets acquire nonzero vacuum expectation values, $L$ is
spontaneously broken, resulting in the appearance of a massless Goldstone boson
called the Majoron which is a decay product of $\nu_\tau$ in this model.
Because
of the interchange symmetry, the Majoron $J$ and its neutral and charged
partners are odd under a discrete $Z_2$ symmetry whereas the other scalar
particles are even. The Higgs potential consisting of only the latter two
scalar
doublets is of the form of Eq.~(1). In the notation of Ref.~3, it can easily be
shown that
\begin{equation}
\lambda_1 = 2 \lambda_0, ~~~ \lambda_2 = \lambda + {1 \over 2} \eta_{12}, ~~~
\lambda_3 = \eta + \zeta + \xi, ~~~ \lambda_4 = - \zeta, ~~~ \lambda_5 = - \xi;
\end{equation}
\begin{equation}
\mu_{12}^2 = 0, ~~~ v_2 = v_0, ~~~ v_1 = \sqrt 2 v_L. \end{equation}
Furthermore
$v_1$ is required to be small, say of order 20 GeV, in this model for various
cosmological and astrophysical reasons\cite{3}. We have thus all the features
necessary for a light $h$ which may have escaped experimental detection. In
addition, the dominant decay of $h$ in this model is into two Majorons ($JJ$)
which interact very weakly and are thus invisible. The signature of $Z
\rightarrow hh \bar {f} f$ is then $Z \rightarrow \bar{f} f$ + missing energy.
This decay mode is free of QCD backgrounds and may be observable if its
branching fraction is greater than 10$^{-6}$.

Suppose we replace $h$ by $J$ and consider the decay of $Z \rightarrow JJ \bar
{f} f$. It appears at first sight that this may have an enhanced branching
fraction, using the same argument as we have presented. Actually, this is not
the case. The reason is that $J$ is a Goldstone boson, so that it can always be
parametrized exponentially, {\it i.e.} as a phase. Hence it has only derivative
couplings in this representation and the analog of diagram (a) does not exist.
In addition, the analog of diagram (b) cannot be enhanced; in the limit of $m_h
= m_H$ it is in fact suppressed by a factor $(k_1 \cdot k_2)/(\lambda_3 +
\lambda_4 + \lambda_5) v_2^2$ relative to that of $Z \rightarrow hh \bar {f}
f$.
In the linear representation where $J$ is treated on the same footing as the
other particles, this suppression manifests itself as a necessary cancellation
between diagrams (a) and (b), and we arrive at the same physical amplitude. In
other words, the $hJJ$ and $HJJ$ couplings are not arbitrary, but are related
to
the other parameters of the model in a completely determined way.

\section{Concluding Remarks}

If there are two or more Higgs doublets, a light neutral $h$ may exist which
has
so far escaped experimental detection. This is not possible in the minimal
supersymmetric standard model, but is natural in any other model where
$\mu_{12}^2 = 0$ and $\lambda_5 \neq 0$, as in the doublet Majoron
model\cite{3}. The process $Z \rightarrow hh \bar {f} f$ is then of interest
because it may be observable for a reasonable $hhH$ coupling if $m_H$ is not
much greater than its current lower bound of 60 GeV. This is especially so if
$h$ decays invisibly, as in the doublet Majoron model. \vspace{0.5in}

\begin{center} {ACKNOWLEDGEMENT}
\end{center}

We thank A. Sopczak for discussions. This work was supported in part by the
U.S.
Department of Energy under Grant No. DE-FG03-94ER40837.

\newpage
\bibliographystyle{unsrt}

\vfill
\newpage
\leftline{{\Large\bf Figure Captions}}

\begin{itemize}
\item[Fig. 1~:] {{The four tree-level diagrams that contribute to $Z\rightarrow
hh\bar {f}f$ in extensions of the standard model with two Higgs doublets. Wavy
lines and solid lines represent Z-bosons and fermion-antifermion pairs,
respectively. Dashed lines represent scalar fields and are labeled as $h$ (the
light Higgs scalar), $H$ (the heavy Higgs scalar), or $A$ (the Higgs
pseudoscalar).}\label{fig1}}

\item[Fig. 2~:] {{ The total branching fraction $B$ for $Z \rightarrow hh
\bar {f} f$ due to the model-independent contribution of Fig.1a alone as a
function of the light Higgs scalar mass
$m_h$.}\label{fig2}}

\item[Fig. 3~:] {{ The total branching fraction $B$ for $Z \rightarrow hh
\bar {f} f$ due to the heavy Higgs scalar $H$ mediated contribution of Fig.1b
alone for various values of
$\eta \equiv |\lambda_3 +
\lambda_4 + \lambda_5|$ (see Eq. (1)), $m_h$, and $m_H$. The decay width of
$H$, due primarilly to $H\rightarrow hh$ and $H\rightarrow \bar {b} b$ (where
we
assume the b-quark Yukawa coupling of the standard model), has been taken into
account. (a): $B$ when $m_h =10$ GeV as a function of $m_H$ for differnt values
of $\eta$. (b): $B$ when $m_h =10$ GeV as a function of $\eta$ for differnt
values of $m_H$. (c): $B$ when $\eta =1$ as a function of $m_h$ for differnt
values of $m_H$.}\label{fig3}}

\end{itemize}

\vfill

\end{document}